\documentclass[12pt]{article} 
\baselineskip 

\usepackage{color}

\usepackage[dvips]{graphicx}
\usepackage{epsfig}

\begin{document}
arXiv 2012
\begin{center}
{\bf{Turbulent Cascades and the $\alpha$ dynamo}}\\

{\em{ J B Taylor}}\\

Radwinter, Wallingford, OX10 9EJ,  UK \\
July 2012
\end{center}

\begin{abstract}
The well known mean-field model of a turbulence driven dynamo is reviewed in relation to Laboratory experiments in which a turbulent cascade is created by a pair of large rotors.   It is argued that in such experiments the
$\alpha$-effect, driving a dynamo field, will be much less than the dissipative $\beta$-effect.   Consequently a mean field dynamo cannot be sustained.   This conclusion is supported by recent measurements of the  $\alpha$ and $\beta$ effects in the Madison Dynamo Experiment.

\end{abstract}

\begin{flushleft}
1 Introduction

The generation of macroscopic magnetic fields in turbulent plasmas is a long standing problem in both astrophysics and in laboratory experiments.   Theoretically, this problem is often addressed through the Mean Field MHD model~\cite{K+R} - in which the emf induced by small scale turbulent fluctuations is averaged over larger scales.   If the turbulence is not reflectionally symmetric, the resulting mean-emf has both a term that amplifies an existing macroscopic magnetic field (the $\alpha$-effect) and a term that dissipates it (the $\beta$-effect).    If the $\alpha$-effect outweighs the $\beta$-effect, amplification of an initial magnetic field may lead to a self sustaining ``mean field" dynamo.    

2 Turbulent Cascade

In laboratory experiments such as the Madison Dynamo Experiment~\cite{Norn 2006} and the VKS Experiment~\cite{VKS 2007}, turbulence is induced by large rotating impellers.   These rotors drive a turbulent cascade of ever smaller eddies that are eventually dissipated by viscosity - in the classic picture of Richardson, Kolmogorov and others.
The intermediate scale eddies form an inertial range in which the energy spectrum $ E(k) =\;<v(k)v^*(k)>\!\!/2 $ depends only on the wave-number $k$ and the rate $ \varepsilon $ at which energy is injected.   Then dimensional considerations lead to the celebrated Kolmogorov spectrum
$ E(k)\sim\varepsilon^{2/3} k^{-5/3}$ .

The inertial range can be regarded~\cite{Bris} as the result of energy  transfer from wave-numbers $< k $ to wave-numbers $>k $, at a rate  determined by a characteristic eddy distortion time $\tau(k)$,  (so that
$\varepsilon \sim kE(k)/ \tau(k) $).   This distortion time is also a measure of the correlation time of an eddy.   It is related~\cite{Kra1971} to the shear across an eddy of size $1/k$ produced by fluctuations of wave-number $<k$,
\begin{equation}1/\tau(k) \sim \left( \int^k k^2 E(k)\;dk\right)^{1/2}
 \sim \varepsilon^{1/3} k^{2/3}
\end{equation}
however the form of $\tau(k)$ is not required in the following analysis.

If the rotors also inject helicity  $ \bf{(v \cdot \nabla \times v)}$ at a rate $\mu$, and this cascades with the same characteristic eddy time $\tau(k)$, then the helicity spectrum $F(k) = \bf{<(v^*(k) \cdot k \times v(k)>}$ is also fixed~\cite{Bris} as $ F(k) \sim \mu\varepsilon^{-1/3}k^{-5/3}$.   However, the  helicity at any scale $k$ cannot exceed $k$ times the energy at that scale, and in the experiments the energy and the helicity are injected at the same (rotor) scale $k_{Rot}$.   Consequently the helicity injection rate $\mu$ must be less than $k_{Rot}$ times the energy injection rate, that is $ \mu < 2 \varepsilon k_{Rot} $.     

3)  Mean emf

To determine the mean emf created by velocity fluctuations in the turbulent cascade, we follow the conventional development~\cite{K+R} based on the MHD eqs for the large scale magnetic field $\bf{B}$,
\begin{equation}
\frac{\partial \bf{B}}{\partial t} = \bf{\nabla \times \xi} +
\bf{\nabla \times ( v \times B)}  +\eta \nabla^2 \bf{B}
\end{equation}
and the small scale magnetic fluctuations $\bf{b}$,
\begin{equation}
\frac{\partial \bf{b}}{\partial t}= \bf{\nabla \times(v \times B)}
+ \eta \nabla^2 \bf{b} + \bf{\nabla \times G}
\end{equation}
Here $\bf{\xi} = <\bf{v} \times \bf{b}> $ is the mean emf we are interested in, and $\bf{G}$ is a ``residual" non-linear term $(\bf{(v\times b) - <v \times b>})$.
It is conventional to neglect this non-linear term.   Then if the plasma resistivity $\eta$ is small, 
\begin{equation} 
{\bf{b}} = \int {\bf{\nabla \times (v \times B)}} dt
\end{equation}
and the mean emf can be written in terms of the velocity fluctuations alone as
\begin{equation}
{\bf{\xi}} =
<{\bf{v}}(x,t)\times \int^t {\bf{\nabla}} \times ({\bf{v}}(x,t') \times {\bf{B}})\; dt'>
\end{equation}
If the turbulence is isotropic this can be reduced to the standard form
\begin{equation}
\bf{\xi} = \alpha \bf{B} - \beta \bf{J}  
\end{equation}
where $\alpha$ is related to the helicity
\begin{equation}
\alpha =
\frac{1}{3}\int <{\bf{v}}(x,t)\cdot\nabla \times{\bf{v}}(x,t-\tau)> \; d\tau
\end{equation}
and $\beta$ is related to the energy
\begin{equation}
\beta = \frac{1}{3}\int <{\bf{v}}(x,t)\cdot {\bf{v}}(x,t-\tau)> \; d\tau
\end{equation}
In terms of the correlation time $\tau(k)$ and the inertial range spectra $E(k)$and $F(k)$,
\begin{equation}
\alpha = \frac{1}{3}\int F(k) \tau(k) \,dk
\sim \mu\varepsilon^{-1/3}\int k^{-5/3}\tau(k) dk
\end{equation}
\begin{equation}
\beta = \frac{2}{3}\int E(k)\tau(k) \,dk
\sim 2 \varepsilon^{2/3} \int k^{-5/3} \tau(k) dk
\end{equation}
so that $\alpha/\beta = \mu/2\varepsilon $ - which we saw earlier is less than 
$< k_{Rot}$.

4 Laboratory Dynamos

According to the expression (Eq(6)) for the mean emf, the power input to any large scale dynamo magnetic field from turbulence is  $ \alpha \bf{J \cdot B}$, while the power lost by this field through turbulent dissipation is $\beta {\bf{J^2}}$.    Consequently, the ratio
\begin{equation} 
\frac{power\; input\;to\;dynamo}{power\; loss\;by\;dynamo} = 
(\frac{\alpha}{\beta}) (\frac{\bf{J\cdot B}}{J^2}) < 
k_{Rot} (\frac{1}{k_{Dyn}})
\end{equation}
where $1/k_{Dyn} \sim B/J $ is the scale length of the dynamo field.

In laboratory experiments (though not in astrophysical plasmas), the energy and helicity injection scale $k_{Rot}$, and scale of the dynamo field $k_{Dyn}$, are similar.  Eq(11) therefore implies that the ratio of power-input to dissipation is of order unity.   Bearing in mind that the input power must also overcome the intrinsic ohmic dissipation this already suggests that creation of a laboratory mean field dynamo is a marginal process.   However, the true situation is much worse.   The inequality
$\mu/2 \varepsilon  < k_{Rot} $ is a very strong one;  equality could be reached only if all fluctuations produced by the rotors were waves of maximal helicity (i.e. all 100\% left or right polarized).   This does not seem  possible, and in any event such a pure state of maximal helicity would not survive non-linear interactions~\cite{Kra1973}.

5 Conclusion

It appears, therefore, that in laboratory dynamo experiments with large rotors, power input from turbulence to the mean magnetic field is likely to be substantially less than the turbulent dissipation - and a mean field dynamo cannot therefore be maintained.   (This conclusion is supported by recent measurements of the mean emf due to plasma fluctuations~\cite{Prep} in the Madison Dynamo Experiment - which showed that the $\alpha$ effect is indeed much smaller than the $\beta$ effect.)

This argument does not, of course, rule out the possibility of a {\em{laminar}} laboratory dynamo, and it leaves open the possibility of a dynamo driven by several small rotors (so that $k_{Rot} >> k_{Dyn}$ in Eq(11)) rather than a single pair of large rotors.   

Also, it does not apply to Astrophysical dynamos which, in the light of the present discussion are seen to differ fundamentally from the laboratory variety.

6 Appendix

The analysis above, which depended on the loosely defined eddy lifetime $\tau(k)$, assumed small plasma resistivity - as seems appropriate for the experiments.   However, in the high resistivity limit a more specific  calculation, which does  not involve $\tau(k)$, leads to similar conclusions. 

In this case we can take Eq(3) in the form
\begin{equation}
(-i\omega + \eta k^2)\bf{b}(\bf{k},\omega) =
i(\bf{k \cdot B})\bf{v}(\bf{k},\omega)
\end{equation}
which again leads to the standard expression $(\alpha \bf{B} - \beta \bf{J})$
for the mean field $\bf{\xi}$, where now~\cite{Moff}
\begin{equation}
\alpha = \frac{\eta}{3}
\int \int \frac{k^2 F(k,\omega)}{\omega^2 + \eta^2k^4}\;dk\;d\omega
\end{equation}
\begin{equation}
\beta =\frac{2\eta}{3}
\int \int \frac{k^2 E(k,\omega)}{\omega^2 + \eta^2k^4}\;dk\;d\omega
\end{equation}
When $ \eta k^2 > \omega $, these become
\begin{equation}
\alpha = \frac{1}{3\eta}\int \frac{ F(k)}{k^2}\;dk
\end{equation}
\begin{equation}
\beta =\frac{2}{3\eta}
\int \frac{ E(k}{k^2}\;dk
\end{equation}
Introducing the inertial range spectra, and noting that 
$\mu < 2 \varepsilon k_{Rot} $  again leads to the conclusion that power input to the mean magnetic field from a turbulent cascade will be much less than the additional turbulent dissipation.

\end{flushleft}


\end{document}